\documentclass[10pt,prd,twocolumn,preprintnumbers,superscriptaddress,showpacs]{revtex4-1}
\usepackage{graphicx,amsmath,amssymb}

\usepackage[
       colorlinks=true,
      filecolor=black,
       anchorcolor=blue,
      linkcolor=blue,
      citecolor=red,
      urlcolor=blue,
       linktocpage=true,
        plainpages=false,
        breaklinks=true,
            pdfstartview=FitH
           ]{hyperref}
\usepackage{verbatim}
\usepackage{yfonts}
\usepackage{color}
\usepackage{ifsym}
\usepackage{lipsum}
\usepackage{hyperref}

\def\nn{\nonumber}

  \newcommand{\<}{\langle\,}

  \renewcommand{\>}{\rangle}
 \def\label{\label}

\renewcommand{\(}{\left(}
\renewcommand{\)}{\right)}
\renewcommand{\[}{\left[}
\renewcommand{\]}{\right]}
\def\e{\varepsilon}

\def\32{{3 \over 2 } }

\begin{document}
\title{Towards Searching for Entangled Photons in the CMB Sky}
\author{Jiunn-Wei Chen}\email{jwc@phys.ntu.edu.tw}
\affiliation{Department of Physics, Center for Theoretical Sciences, and Leung Center for Cosmology and Particle Astrophysics, National Taiwan University, Taipei 10617, Taiwan}
\affiliation{Center for Theoretical Physics, Massachusetts Institute of Technology, Cambridge, Massachusetts 02139, USA}
\author{Shou-Huang Dai}\email{shdai.hep@gmail.com}
\affiliation{Center for General Education, Southern Taiwan University of Science and Technology, Tainan 71005, Taiwan}

\author{Debaprasad Maity}\email{debu.imsc@gmail.com}
\affiliation{Department of Physics, Indian Institute of Technology, Guwahati, India}

\author{Sichun Sun}\email{sichunssun@gmail.com}
\affiliation{Department of Physics, Center for Theoretical Sciences, and Leung Center for Cosmology and Particle Astrophysics, National Taiwan University, Taipei 10617, Taiwan}
\affiliation{Department of Physics, Sapienza University of Rome, Rome I-00185, Italy}

\author{Yun-Long Zhang}\email{yunlong.zhang@apctp.org}
%

\affiliation{Asia Pacific Center for Theoretical Physics, Pohang 790-784, Korea}
\affiliation{Center for Quantum Spacetime, Sogang University, Seoul 121-742, Korea}
\affiliation{Yukawa Institute for Theoretical Physics, Kyoto University, Kyoto 606-8502, Japan}

\date{\small February 25, 2019}

\preprint{MIT-CTP/4870}
\pacs{98.70.Vc; 42.50.Xa; 03.65.Ud 
~~~ arXiv: \href{https://arxiv.org/abs/1701.03437 }{1701.03437 }
~~ DOI:  \href{https://doi.org/10.1103/PhysRevD.99.023507}{10.1103/PhysRevD.99.023507} }



\begin{abstract}
We explore the possibility of detecting entangled photon pairs from cosmic microwave background or other cosmological sources coming from two patches of the sky. The measurements use two detectors with  different photon polarizer directions. When two photon sources are separated by a large angle relative to the earth, such that each detector has only one photon source in its field of view, a null test of unentangled photons can be performed. The deviation from this unentangled background is, in principle, the signature of photon entanglement. To confirm whether the deviation is consistent with entangled photons, we derive a photon polarization correlation to compare with, similar to that in a Bell inequality measurement. However, since photon coincidence measurement cannot be used to discriminate unentangled cosmic photons,  it is unlikely that the correlation expectation value alone can violate Bell inequality to provide the signature for entanglement. 


\end{abstract}

\maketitle

\allowdisplaybreaks

 \section{Introduction}

 The cosmological microwave background and the other sources of the comic photon can reveal much information of the early universe. Especially in an inflationary scenario, quantum fluctuations, which are usually of atomic scales, can be amplified to cosmological scales(see e.g. \cite{Linde:2005ht,Guth:1980zm} ),
and their traces can be seen from the cosmological microwave background (CMB) \cite{Ade:2015xua,Array:2015xqh}. 
Some alternatives to inflationary cosmology can also produce similar fluctuations that are classical in nature. 
In one class of studies \cite{Hartle:1992as,Maldacena:2015bha,Choudhury:2016cso,Polarski:1995jg,Kiefer:1999sj,Kiefer:2008ku,Martin:2007bw,Martin:2015qta,Martin:2016tbd}, entanglement, which can be demonstrated by the violation of Bell inequality, is used to show that whether the CMB fluctuations come from classical theories.


In the standard ``inflationary cosmology'', the density fluctuations that we observe in the universe today are thought to come from primordial quantum fluctuations.
However, it is still a question that how to detect any of those early universe quantum information directly, due to the facts that those part of the information may have already been washed away today \cite{Nambu:2013rta}. 
It is difficult to discuss how quantum entanglement can survive during the inflation or other types of cosmic expansion, such that two particles from two patches of the sky can still be entangled.
This involves how the quantum entanglement is produced, transferred and how part of the entanglement is destroyed by decoherence during the evolution of the universe, where the possibility indeed exists in principle \cite{Nambu:2013rta,Collins:2016ahj,MartinMartinez:2012sg}.
On the other hand, there are also some alternative models, such as the ``wormhole cosmology"  in \cite{Hochberg:1992du} which may provide more entanglement in the universe with a new mechanism.

Besides from this type of entanglement of CMB photons, it is known that a parent particle decays into two or more particles which forms an entangled quantum state as a consequence of conservation laws in the decay process. For example, the decay product of unknown matter, eg. dark matter, or even the primordial black holes. The final diphoton decay products is a likely source of entangled cosmological photons \cite{Lello:2013bva, Valentini:2006yj}.


In this paper,  we ask ourselves a rather elementary and model independent question: If there are entangled photon pairs coming from two different patches of the CMB sky, how do we tell that they are entangled? And if we do not know how large the signal and background are {\it a priori}, how do we isolate the signal from the background?

Our starting point is the condition that if the two photons are {\it unentangled}, then the two-photon density matrix $\pi$ can be written as a direct product of two one-photon density matrix elements:   
\begin{align} \label{null}
 \pi=  \pi_1 \otimes \pi_2 ,
\end{align}
where $\pi_{i}\, (i= 1,2)$  is the one-photon density matrix element from source $i$ and is computable. Therefore, one can perform a {\it null test} to see whether the two photons from the two sources are consistent with unentangled photons. The deviation from this unentangled background is, in principle, the signature of photon entanglement. However, to further check whether the deviation is consistent with entangled photons, we need to study its characteristics.


We consider two scenarios. 
The first scenario is shown in Fig.{~}\ref{hbtGeometry} with the two photons, which might be causally connected in the far past, coming from two very close patches of the sky, such that both sources are in the fields of view of both detectors. The second scenario is shown in Fig.{~}\ref{hbtGeometry2} with two sources in a relatively large angle to the earth such that each detector has one source in its field of view. The main difference is that the interference between different photon paths only happens in the first scenario but not in the second one. It turns out that the second scenario allows simpler isolation of the unentangled photon background from the possible entangled photon signal by adjusting the orientation of the polarizers. 
 We have not attempted to estimate the size of the signal, which depends on the particular model. We will leave it for future studies.

In the following section \ref{ReviewBell}, we review the lesson we learn from the Bell inequality for two photon systems and explain why it cannot be directly applied in CMB photons.
In section \ref{CMB}, we explore the possibility of detecting entangled photon pairs from small angle and large angle, respectively,  in two scenarios.
We present the summary and conclusion in section \ref{Con}.




\section{Lessons from Bell inequality for two photon systems}
\label{ReviewBell}

In the helicity basis, a photon state with $\pm\,\hbar$ helicity can be denoted as $ |\e_{\pm}\>\propto |\e_1\>\pm i |\e_2\>$, with $\e_1$ and $\e_2$ the two orthogonal linear polarizations perpendicular to the direction of propagation. Then there are four independent
two photons spin eigen wave functions: $ |\e_{+}{\>}\otimes{|}\e_{+}\>$,  $|\e_{-}{\>}\otimes{|}\e_{-}\>$, $|\e_{+}{\>}\otimes{|}\e_{-}\>+|\e_{-}{\>}\otimes{|}\e_{+}\> $ and $|\e_{+}{\>}\otimes{|}\e_{-}\>-|\e_{-}{\>}\otimes{|}\e_{+}\> $. The latter two are entangled states which can be rewritten in linear polarization basis as
\begin{align}\label{WF}
|\psi_1 \rangle &= \frac{1}{\sqrt{2}}\[ |\e_1\> \otimes |\e_1\> + |\e_2\> \otimes |\e_2\> \],\nn\\
|\psi_2 \rangle &= \frac{1}{\sqrt{2}}\[ |\e_1\> \otimes |\e_2\> - |\e_2\> \otimes |\e_1\> \].
\end{align}


 Suppose the measurement of photons takes place at two spatially separated locations labelled by Alice (abbreviated by A) and Bob (abbreviated by B). $\Pi_A$ and $\Pi_{A'}$ are projections applied at detector A, while $\Pi_B$ and $\Pi_{B'}$ at detector B. Consider the combination spin-spin operators proposed by CHSH \cite{CHSH}
\begin{align} \label{CHSHp}
C &=  {\Pi_A\Pi_B} +  \Pi_{A'}  \Pi_B + {\Pi_A \Pi_{B'}} - {\Pi_{A'}  \Pi_{B'}} , \nn\\ 
\Pi_I &\equiv | n_I\> \< n_I |-| n_{I_{\perp}}\> \< n_{I_{\perp}}| \,,\quad ( I=A,B,A',B' ) .
\end{align}
In which the projection operator $\Pi_A$ gives a value $+1$ when a photon with polarization in the $\vec{n}_A$ direction is detected, and $-1$ when a photon with polarization perpendicular to $\vec{n}_A$ (denoted as $\vec{n}_{A_{\perp}}$) is measured. 
In other words, each photon registers a $+1$ or $-1$ at the detector with polarizer $\Pi_I$ ($I=A,B,A',B'$). Therefore, if $\{\Pi_A,\Pi_{A'}\}$ register $\{+1,+1\}$ or $\{-1,-1\}$, then the last two terms in Eq.(\ref{CHSHp}) cancel. If $\{\Pi_A, \Pi_{A'}\}$ register $\{+1,-1\}$ or $\{-1,+1\}$, then the first two terms in Eq.(\ref{CHSHp}) cancel. This leads to the Bell inequality $ |\langle C \rangle | \leq 2$ appropriate for a local classical hidden variable theory.

On the other hand, in quantum mechanics, the Bell inequality can be violated~\cite{Bell:1964kc,QP}, and the expectation value of $C$ can be bigger, satisfying $|\langle C \rangle |\leq 2\sqrt{2}$ \cite{Cirelson}, as $C^2   =   4 {\mathbb{I}} -   [ \Pi_A,\Pi_{A'}][\Pi_B,\Pi_{B'}]  =4(1+\sin 2 \theta_{AA'}\sin 2 \theta_{BB'}) {\mathbb{I}} $,
 with $\theta_{AB}$ the angle between $\vec{n}_A$ and $\vec{n}_B$.

Since
\begin{align}\label{thetaAB}
\langle \psi_{i}|{\Pi_A \Pi_B}|\psi_{i}\rangle= (-1)^{i+1} \cos 2\theta_{AB} ,
\end{align}
for $i=1,2$, then for both $|\psi_1 \rangle$ and $|\psi_2 \rangle$, the quantum mechanical bound can be saturated by choosing
\begin{align}
\theta_{AB}=\theta_{AB'}=\theta_{A'B} = \pi/8,\quad \theta_{A'B'}= 3\pi/8.
\end{align}
For unentangled pure states, $\langle \psi_{i}|{\Pi_A \Pi_B}|\psi_{i}\rangle=0$. 

Therefore, if an observation can be performed that pre-excludes unentangled states and show that the expectation value 
$2 < |\langle C \rangle |\leq 2\sqrt{2}$ is achieved, then entanglement between two photons can be established. To exclude the unentangled states, one can control the photon source to make sure most of the photon pairs produced are entangled photons, and then set up coincidence measurement to make sure the detected photons are produced from the source at the same time. Although this condition can be met for controlled experiments in laboratories, it is not the case for CMB photons because coincidence measurement can not guarantee that two photons detected were produced at the same time. Therefore, for CMB photons, there will be lots of unentangled photon background that can be hardly removed. These unentangled photons give the dominant contribution to $ \langle C \rangle$ and lead to  $ |\langle C \rangle | \leq 2$. This renders it unpractical to perform a Bell Inequality type experiment for CMB photons. However, after we perform the null test using Eq.(\ref{null}), we can still use the angular dependence for entangled photon pairs in Eq.(\ref{thetaAB}) to check whether the deviation from the null result is consistent with entangled photon pairs.


\section{Detection of CMB photons}\label{CMB}

In this section, we follow the procedure to first study the null test of unentangled photons using Eq.(\ref{null}), then we study if any deviation from the unentangled result is found, then what kind of signature is consistent with entangled photon pairs.


\subsection{Scenario I: Sources From Small Angles --- Interference}

\begin{figure}[t]
\begin{center}
\includegraphics[scale=0.35]{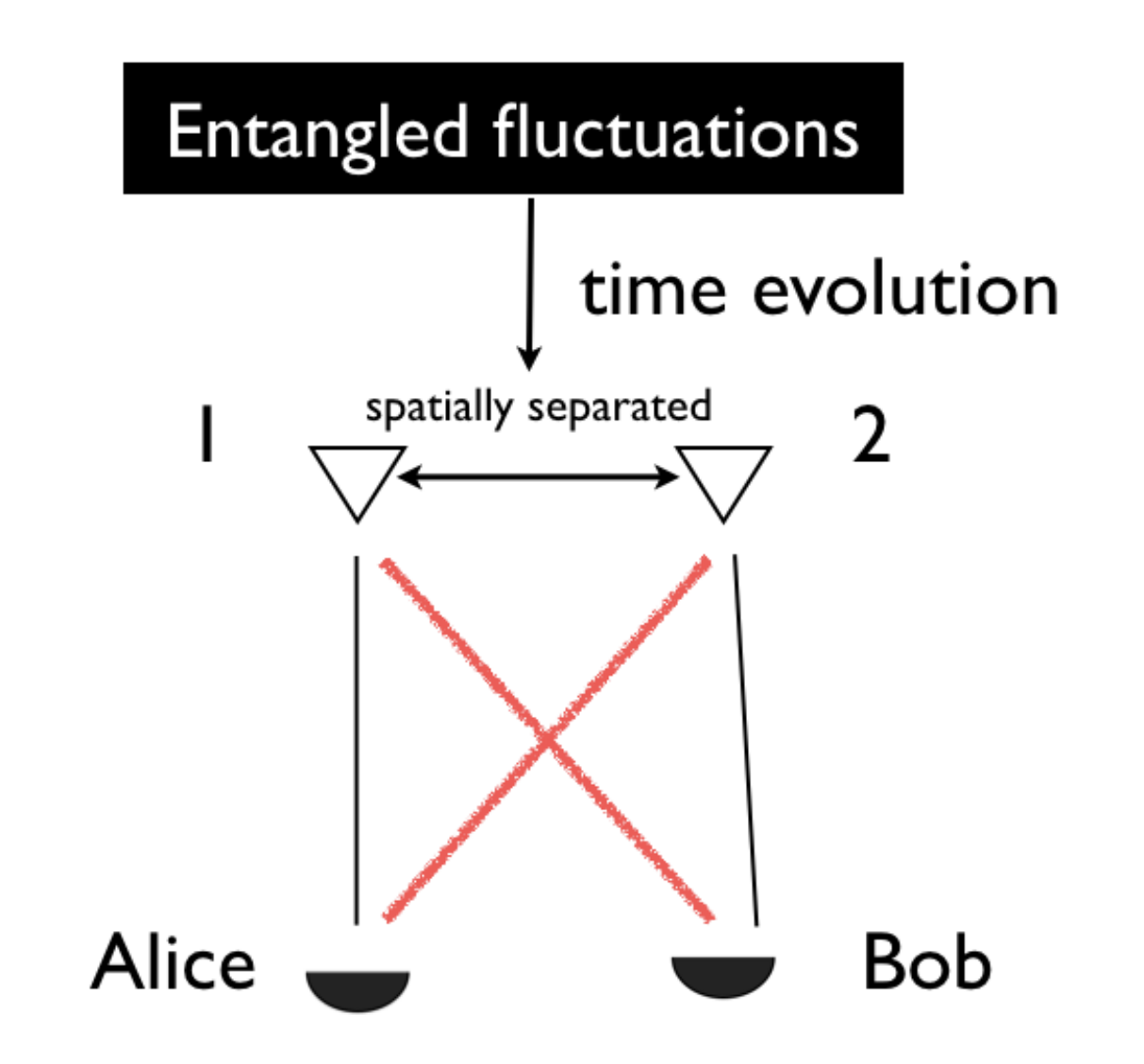}
\caption{Assuming a pair of entangled photons were sent to two regions 1 \& 2 in the sky after inflation,
then the two photons are detected by two detectors Alice and Bob. Since both 1 \& 2 are in the fields of view for the two detectors, the photons can be detected by going through the black or red paths. Correlations of photon polarizations are measured to detect entanglement.
\label{hbtGeometry} }
\end{center}
\end{figure}

The first scenario is when the two photons, which might be causally connected in the far past, are coming from two patches of the sky (called sources 1 and 2) that are close enough to each others, such that sources 1 and 2 are both in the fields of view of the detectors Alice (A) and Bob (B). This is a Hanbury-Brown-Twiss (HBT) intensity interferometer \cite{HanburyBrown:1956bqd,Brown:1956zza, Baym:1997ce} as shown in Fig.{~}\ref{hbtGeometry}.

In HBT, we consider the process with two photons emitted from source 1 and 2 in the initial state and those two photons are detected by detectors A and B in the final state. The propagation of photons is described by Feynman's path integral.
The amplitude can be written as
\begin{align}
\mathcal{A}=D_{1A}D_{2B}  +   D_{2A}D_{1B} ,
\end{align}
where  $D_{1A}$ denotes the propagator from source 1 to detector $A$, and so on. The two different terms correspond to two different paths shown in Fig.~\ref{hbtGeometry}, and higher order loop diagrams are neglected.  Then the transition probability is proportional to
\begin{align}\label{HBT}
|\mathcal{A}|^2=
&  | D_{1A} |^2   | D_{2B}|^2 +  | D_{2A}|^2   |D_{1B} |^2  \nn \\
& + 2  {\rm Re}  D_{1A}D_{2B}D_{2A}^*D_{1B}^*.
\end{align}
Since the photon from each source appears both in a propagator and a conjugate one, the random phases associated with each of the sources 1 and 2 cancel in the interference term. Therefore, the coherent source is not required. For simplicity, we assume the two photons are produced and emitted at the same time, thus the interference term depends only on the relative phase factor arising from the geometry of the setup. The longer the baseline (i.e. the distance between the two detectors), the better the resolution for the interferometry. One can generalize the analysis to cases with the two sources coming from an extended object such as a star or the CMB sky.


\subsubsection{The Null Test for Unentangled Photons}

The possible source of unentangled photon background includes:
Type (a), two uncorrelated photons coming from two different sources;
Type (b), two uncorrelated photons coming from the same source.  Type (b) background can be considered as a special case of Type (a). Hence we will consider Type (a) first.


Consider the two unentangled photons are coming from sources 1 and 2 with one photon density matrix $ \pi_1$ and $ \pi_2$ 
satisfying Eq.(\ref{null}).
$\pi_i$ at source $i$ with net polarization $\alpha_i$ in the $\vec{n}_i$ direction can be written by
\begin{align} \label{bg}
\pi_i=\frac{1}{2+2\alpha_i}\Big{[}(1+2\alpha_i)|n_i\>\< n_i|+|n_{i_{\perp}}\>\< n_{i_{\perp}}|\Big{]},
\end{align}
where $\vec{n}_{i_{\perp}}$ is perpendicular to $\vec{n}_i$ and $i=1,2$.

For null test of unentangled background, the total probability for detectors A and B each detects one photon is \cite{Wilczek2015}
\begin{align}\label{direct}
&{\rm Tr} \(\Pi_A  \pi_1\)   {\rm Tr}  \(\Pi_B  \pi_2\) |D_{1A}|^2 |D_{2B}|^2 \,\nn\\
+&{\rm Tr} \(\Pi_A  \pi_2\)  {\rm Tr} \( \Pi_B  \pi_1\) |D_{2A}|^2 |D_{1B}|^2\,\nn\\
+& {\rm Tr} \( \Pi_A \pi_1 \Pi_B \pi_2 \) D_{1A}D_{2B}D_{2A}^*D_{1B}^*   \,\nn\\
+& {\rm Tr} \(\Pi_A \pi_2 \Pi_B \pi_1 \) D_{1A}^*D_{2B}^*D_{2A}D_{1B}.
\end{align}
Then, by using Eq.\eqref{CHSHp} and Eq.\eqref{bg},  we have the formula of the first line in the total background \eqref{direct}
\begin{align}\label{bg1}
{\rm Tr} \(\Pi_A  \pi_1\)   {\rm Tr}  \(\Pi_B  \pi_2\)=\frac{\alpha_1 \alpha_2 \cos 2 \theta_{An_1}\cos 2 \theta_{Bn_2}}{(1+\alpha_1)(1+\alpha_2)},
\end{align}
where  $\theta_{An_i}$ is the angle between the orientation of the polarizer at detector A and $\vec{n}_i$. The part of $\vec{n}_{i \perp}$ has been converted into $\vec{n}_i$ by using trigonometry formulas. ${\rm Tr} \(\Pi_A  \pi_2\)  {\rm Tr} \( \Pi_B  \pi_1\)$ in the background \eqref{direct} can be obtained analogously by interchanging $1$ and $2$. The geometrical phase dependent interference terms in the background \eqref{direct} yield
\begin{align}\label{bg2}
{\rm Tr} \( \Pi_A \pi_1 \Pi_B \pi_2 \) ={\rm Tr} \(\Pi_A \pi_2 \Pi_B \pi_1 \) \propto \cos 2 \theta_{AB} ,
\end{align}
up to $\mathcal{O}(\alpha_i)$ corrections. When we consider the Type (b) background, we can just set $(1,2)$ to  $(1,1)$ or  $(2,2)$. However, their relative weights are in general different. 

\subsubsection{Consistency Check for Entanglement}
Suppose the two photons emitted from the sources 1 and 2 are entangled to form the $|\psi_{1} \rangle$ state of Eq.(\ref{WF}), then after propagating to the detectors, the photon state becomes
\begin{align}
|\Psi_{1} \rangle&= 
\frac{1}{\sqrt{2}}
\Big( D_{1A} |\e_1\> \otimes D_{2B} |\e_1\>  +D_{1A} |\e_2\> \otimes D_{2B} |\e_2\>  \nn\\
&\quad\quad~ +  D_{2A} |\e_1\> \otimes D_{1B} |\e_1\>  +D_{2A} |\e_2\> \otimes D_{1B} |\e_2\>  \Big)\nn\\
&= 
|\psi_{1} \rangle \left(D_{1A} D_{2B} +D_{2A} D_{1B}\right),
\end{align}
where the spin wave function is not changed during the propagation and the propagation is not changed by the spin.
The same factorization happens for $|\Psi_{2}\>$ as well.
As a result, the expectation value of the CHSH quantity reads
\begin{align} \label{vevce}
\langle \Psi_{i}|\Pi_A \Pi_B|\Psi_{i} \rangle =
\langle \psi_{i}|\Pi_A \Pi_B|\psi_{i} \rangle \left|D_{1A} D_{2B} +D_{2A} D_{1B}\right|^2,
\end{align}
with the spin-spin correlation $\langle \psi_{i}|\Pi_A \Pi_B|\psi_{i} \rangle$ and the geometric phase contribution $\left|D_{1A} D_{2B} +D_{2A} D_{1B}\right|^2$ completely factorized in the measurement.
Therefore, entanglement between two  CMB  photons emitted from two sources 1 and 2 (from two patches of the sky) yields the $\cos 2 \theta_{AB}$ dependence of
the polarization correlation between two detectors A and B (see Eq.(\ref{thetaAB})):
\begin{align} \label{signal1}
\langle \Psi_{i}|\Pi_A \Pi_B|\Psi_{i} \rangle \propto \cos 2 \theta_{AB},
\end{align}
with $i=1,2$. This can be achieved by changing $\theta_{AB}$. Note that this conclusion also applies to the special case where the two photons are coming from the same source.

From the above discussion, we see the geometrical phase dependent background of Eq. (\ref{bg2}) has the same angular dependence as the entanglement ``signal'' of Eq. (\ref{signal1}). This means unless one can compute the background to very high accuracy, one cannot
isolate the signal from the background using this geometrical phase dependent term. The geometrical phase independent terms, e.g. Eq. (\ref{bg1}), seem to have different angular dependence to ``signal'' of Eq. (\ref{signal1}). 

However, if we consider another background from two photons coming from the same source, which can be computed by replacing the $(1,2)$ indices of Eq. (\ref{direct}) by $(1,1)$ and $(2,2)$, then all the background terms do not depend on the geometrical phase, and the same angular dependence as the signal can appear through Eq.(\ref{bg2}).
Therefore, separating the signal from the background is challenging. Fortunately, the situation is simpler when we consider another scenario in the next section.

The set up discussed in this section is also discussed in Ref. \cite{Wilczek2015}, where it was pointed out that deviation from the unentangled photons with density matrix in Eq. (\ref{null}) can be used as a signal for entanglement.



\subsection{Scenario II: Sources From Large Angle --- \\ No Interference}

\begin{figure}[t]
\begin{center}
\includegraphics[scale=0.35]{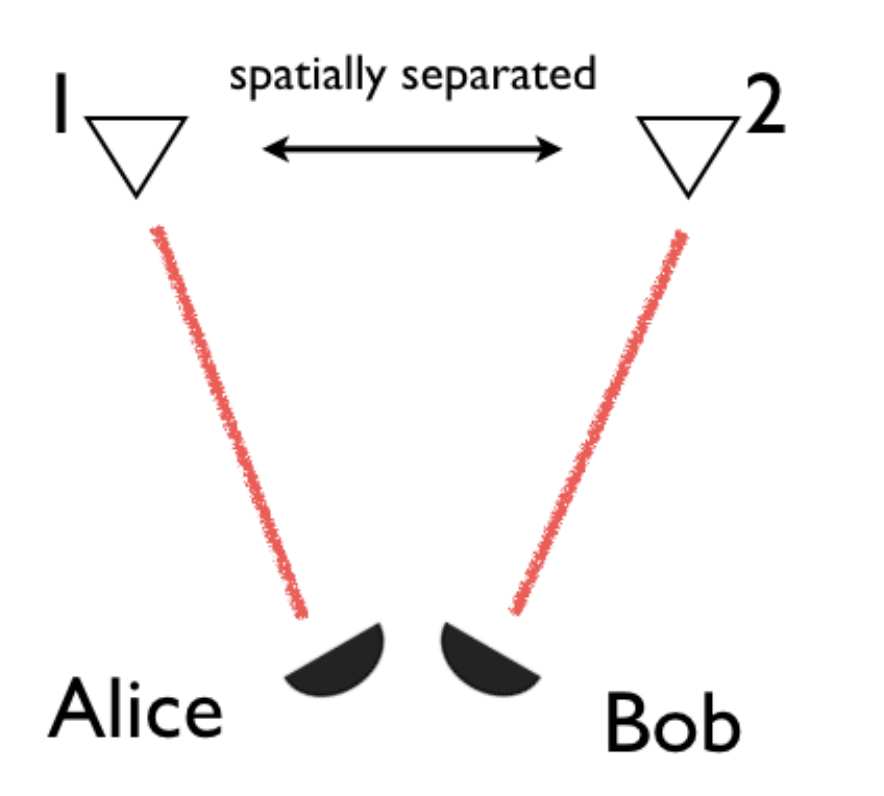}
\caption{This is similar to the set up of Fig.\ref{hbtGeometry}, except 1 is in the field of view of Alice only  and 2 is only in the field of view of Bob. The photons can only be detected after going via one path. Hence there is no interference terms.
\label{hbtGeometry2} }
\end{center}
\end{figure}

When sources 1 and 2 are far away such that 1 is only in the field of view of A and 2 is only in the field of view of B, then there is no interference between different paths, as shown in Fig.{~}\ref{hbtGeometry2}.
We can simply remove one of the paths in Scenario I by setting $D_{2A}=D_{1B}=0$ for the null test of unentangled background from Eq.(\ref{direct}) and double check the entangled signal from Eq. (\ref{vevce}).

\subsubsection{The Null Test for Unentangled Photons}

For null test of unentangled background in Scenario II, we only need to 
 consider the first line in eq.\eqref{direct} with the same pre-factor
 ${\rm Tr} \(\Pi_A  \pi_1\)   {\rm Tr}  \(\Pi_B  \pi_2\)$, whose angular dependence is computed in Eq.(\ref{bg1}) already:
\begin{align}
{\rm Tr} \(\Pi_A  \pi_1\)   {\rm Tr}  \(\Pi_B  \pi_2\) \propto  \cos 2 \theta_{An_1}\cos 2 \theta_{Bn_2}.
\label{eq15}
\end{align}
The background has different angular dependence to the signal, $\cos 2 \theta_{AB}$. For example, one could have chosen the direction of polarizer $A$ such that $\cos 2 \theta_{An_1}=0$ and the background vanishes, then one can subsequently change $\theta_{AB}$ to look for the $\cos 2 \theta_{AB}$ signal. This set up does not have the complication of scenario I where the background is a combination of several terms in \eqref{bg1} and \eqref{bg2}, which makes the isolation of signal from background complicated.

\subsubsection{Consistency Check for Entanglement}

To double check the entangled signal in Scenario II, we have
\begin{align}
\langle \Psi_{i}|\Pi_A \Pi_B|\Psi_{i} \rangle =
\langle \psi_{i}|\Pi_A \Pi_B|\psi_{i} \rangle \left|D_{1A} D_{2B} \right|^2 \propto \cos 2 \theta_{AB}.
\label{eq16}
\end{align}
The entanglement signal still has the $\cos 2 \theta_{AB}$ dependence. By adjusting the detector angles and observing the angular dependence functions, we can distinguish signal in \eqref{eq16} from the background in \eqref{eq15}.


The setting of this large angle scenario is similar to a set up in Refs. \cite{Gallicchio:2013iva,Handsteiner:2016ulx} where lights from two distant sources separated by a large angle are used to determine the polarization directions of the two detectors. This allows pushing back the time for the possible ``freedom of choice" loophole to happen all the way to the time when the two distant light sources were in contact.

\section{Summary and discussion}\label{Con}

We have explored the possibility of detecting entangled photon pairs from CMB or other cosmological sources coming from two patches of the sky. The measurement uses two detectors with photon polarizers in different directions. When two photon sources are separated by a large angle relative to the earth, such that each detector has only one photon source in its field of view, then the signal for entanglement can be separated from the background by changing the polarizer directions. In some special choice of the polarizer directions, the leading uncorrelated photon background can be completely blocked. When the angle between two photon sources is small enough such that both sources are in the fields of view of both detectors, then the background becomes more complicated with several terms, with different angular dependence and different weights. In all, the large angle scenario is preferred.

One question we do not discuss here in this paper is the different sources for this type of entanglement. Some of the possibilities are: two entangled particles in the casually disconnected regions are connected by a wormhole \cite{Maldacena:2013xja,Jensen:2013ora,Chen:2016xqz,Hochberg:1992du}, or they are the decay products of dark matter candidates or some extensive cosmic objects.  The particle production towards the end of inflation or later by the inflaton may also be a source for these entangled pairs. Previous studies for particle productions only showed the possible signatures in CMB scalar or tensor mode correlations \cite{Cook:2011hg,Barnaby:2009dd}. In fact, those particles produced are diluted away by cosmic expansion and chances to detect them are small. This entanglement between the CMB photon polarizations is different yet related to the old question that whether the origin of cosmic fluctuations is quantum or classical \cite{Martin:2007bw,Martin:2015qta,Martin:2016tbd}, although exact connections require further studies.

Our work is only a first step towards the detection of entangled photons in the sky, and the result of our initial study is encouraging. A critical question is how large the estimated signal is compared with the current detector sensitivity and what would be possible sources of entangled photons. 
More recent studies of the intensity-intensity correlations provide the similar method to investigate the nonclassical nature of photons, see e.g. \cite{Martin:2017zxs, Kanno:2017dci, Kanno:2018cuk}. 
We also hope this proposal can be implemented in the current observations like BICEP/Keck Array in \cite{Array:2015xqh}.
 
Notice that in \cite{MartinMartinez:2012sg}, it was pointed out that even if the quantum entanglement in early universe is hard to be observed through the CMB photons, such signals in the cosmic neutrino background (CNB) are also suggested. Although the CNB is still too weak to be detected now, our proposal in this work can also be generalized to the entangled neutrinos and it is interesting to leave it for a future study.

\subsection*{Acknowledgements}
This work is inspired by a lecture of Frank Wilczek on Entanglement Enabled Intensity Interferometry.
We also thank Dave Kaiser for interesting discussions.
This work is partially supported by the MOST, NTU-CTS and the NTU-CASTS of Taiwan.
J.W.Chen is partially supported by MIT MISTI program and the Kenda Foundation.
S. Sun thanks MIUR (Italy) for partial support under Contract No. PRIN 2015P5SBHT and partial support from ERC Ideas Advanced Grant No. 267985 ``DaMeSyFl''. Y. L. Zhang thanks the support from APCTP and CQUeST, as well as Grant-in-Aid for international research fellow (No. 18F18315) of JSPS.







\end{document}